\documentstyle[fleqn,espcrc2,epsf]{article}
\title{
$I=2$ Pion Scattering Length and Phase Shift with Wilson Fermions
\thanks{presented by N. Ishizuka}
}
\author{
CP-PACS Collaboration : 
S.~Aoki
\address{
Institute of Physics,
University of Tsukuba,
Tsukuba, Ibaraki 305-8571, Japan
},
R.~Burkhalter
\address{
Center for Computational Physics, 
University of Tsukuba,
Tsukuba, Ibaraki 305-8577, Japan
},
M.~Fukugita
\address{
Institute for Cosmic Ray Research,
University of Tokyo,
Kashiwa 277-8582, Japan
},
S.~Hashimoto
\address{
High Energy Accelerator Research Organization(KEK),
Tsukuba, Ibaraki 305-0801, Japan
},
N.~Ishizuka$^{\rm ~a,b}$,
Y.~Iwasaki$^{\rm ~a,b}$,
K.~Kanaya$^{\rm ~a}$,
T.~Kaneko$^{\rm ~d}$,
Y.~Kuramashi$^{\rm ~d}$,
V.~Lesk$^{\rm ~b}$,
M.~Okawa$^{\rm ~d}$,
Y.~Taniguchi$^{\rm ~a}$,
A.~Ukawa$^{\rm ~a,b}$ and
T.~Yoshi\'e$^{\rm ~a,b}$ 
}
\begin{document}
%
%
\begin{abstract}
We present preliminary results of
scattering length and phase shift for I=2 S-wave $\pi\pi$ system 
with the Wilson fermions in the quenched approximation.
The finite size method presented by L\"uscher is employed, 
and calculations are carried out at $\beta=5.9$ on a $24^3\times 60$ and 
$32^3\times 60$ lattice.
\end{abstract}
\maketitle
%
%
\section{ Introduction }
Lattice calculations of scattering lengths and phase shifts 
for the pion system is an important step for understanding 
of strong interactions beyond the hadron mass spectrum.

There are already several studies of scattering lengths in literature.
For the $I=2$ process calculations has been carried out with the 
staggered~\cite{SGK,Kuramashi} and the Wilson fermion 
actions~\cite{Kuramashi,GPS,JLQCD}.
A recent work by JLQCD \cite{JLQCD} carried out the calculation at 
several lattice cutoffs and obtained the value in the continuum limit.
For the $I=0$ process, which is difficult
due to the presence of disconnected contributions,
a pioneering attempt has been made in Ref.~\cite{Kuramashi}.

For the scattering phase shift, in contrast, 
there is only one calculation for $I=2$~\cite{Fiebig}, 
and the data are still too noisy to obtain reliable results.

In this article we report on a high statistics calculation
of the $I=2$ S-wave pion scattering length and phase shift.
We work in quenched lattice QCD employing the standard plaquette 
action for gluons and the Wilson fermion action for quarks.
Simulations are made at $\beta = 5.9$ using 200 and 173 configurations 
on a $24^3 \times 60$ and $32^3 \times 60$ lattice.
Quark propagators are solved with the Dirichlet boundary condition 
imposed in the time direction and the periodic boundary condition in the 
space directions. 
Quark masses are chosen to be the same as in the previous study
of quenched hadron spectroscopy by CP-PACS~\cite{CP-PACS.LHM}
({\it i.e.}, $m_\pi/m_\rho = 0.491, 0.593, 0.692, 0.752$ ) 
for both lattice sizes.
%
%
\vspace{-0.2cm}
\section{ Method }
The energy eigenvalue $W_p$ of an S-wave $\pi\pi$ system
with momentum $\vec{p}$ and $-\vec{p}$ 
in a finite periodic box of a size $L^3$
is shifted from twice the pion energy $2\cdot E_p$ by finite-size effects.
L\"uscher derived a relation between the energy shift 
$\Delta W_p = W_p - 2\cdot E_p$
and the scattering phase shift $\delta(p)$, 
which takes the form \cite{Lusher}
\begin{equation}
\nu_n \frac{ \tan \delta (p) }{ \pi L p } = - x_p - A_n x_p^2 - B_n x_p^3 + O(x_p^4)
\label{Lusher.eq}
\end{equation}
where
$p^2 = n \cdot (2\pi/L)^2$, ($n = 0, 1, \cdots 6$),
$x_p = \Delta W_p \cdot (2 E_p L^2) / (16\pi^2) = O(1/ L)$, 
and $\nu_n$, $A_n$, and $B_n$ are geometrical constants.
The scattering length is given by $a_0 = ( \tan \delta(p) / p )_{p \to 0}$. 

In order to obtain the energy eigenvalue $W_p$
we construct $\pi\pi$ 4-point functions 
$
  G_{pk}(t) = \langle 0 | \Omega_p ( t ) \Omega_k ( 0 ) | 0 \rangle
$ .
Here $\Omega_p( t )$ is an interpolating field for the S-wave $\pi\pi$ system 
at time $t$ given by 
$  \Omega_p( t ) = \sum_{R}
                      \pi(  R(\vec{p}), t )
                      \pi( -R(\vec{p}), t )  $ 
where $R$ is an element of the cubic group. 
In numerical calculations
we construct the source operator $\Omega_k( 0 )$
using the noisy source method with U(1) random numbers.
%
%
\begin{figure}[t]
\vspace*{-0.6cm}
\centerline{\epsfxsize=7.0cm \epsfbox{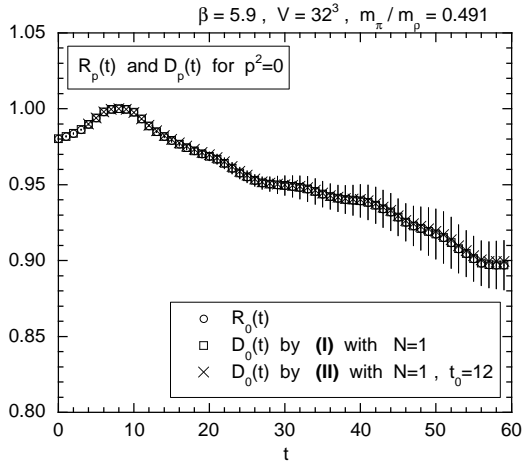}}
\vspace*{-1.0cm}
\caption{\label{Fig1.fig}
$R_p(t)$ and $D_p(t)$ for $p^2=0$.
}
\vspace*{-0.6cm}
\end{figure}
%
%

Since the 4-point function $G_{pk}(t)$ contains many exponential terms
due to the Maiani-Testa no-go theorem~\cite{MT-nogo},
the extraction of energy eigenvalues from $G_{pk}(t)$ is non-trivial. 
The following two methods are known~\cite{Lusher-Wolf} :
\hfill\break
(I)
Diagonalization of $G_{pk}(t)$ at each $t$, 
where the momenta $p$ and $k$ are regarded as matrix indices.
The eigenvalues take the form
$\lambda_p (t) = C_p \exp( - W_p t ) \cdot [ 1 + O( \exp ( - \Delta_p t ) ]$,
where $C_p > 0$ and $\Delta_p$ is the distance of $W_p$ from the other 
eigenvalues $W_k$.
In this method the fitting range of $t$ should be taken large 
so that the $O( \exp ( - \Delta_p t ))$ terms can be ignored.
\hfill\break
(II)
Diagonalization of 
$M_{pk} (t_0,t)$ $\equiv$ $[ G(t_0)^{-1/2}$ $G(t) G(t_0)^{-1/2}]_{pk}$
at each $t$ where $t_0$ is fixed at some small value.
The eigenvalues are
given by $\lambda_p (t) = \exp( - W_p (t-t_0) )$
where $O( \exp ( - \Delta_p t ))$ terms are absent.

In the actual diagonalization we have to cut off the set of momenta.
Here we expect that the components of $G_{pk}(t)$ or $M_{pk}(t_0,t)$ for 
$p,k \leq q$ are dominant for the eigenvalue $\lambda_q(t)$ in large $t$ 
region, while the components $p,k > q$ are less important.
With this expectation, using both methods we calculate the energy eigenvalue
for the ground state ($n=0$) and the first excited state ($n=1$) for $V=24^3$,
and also that of the second excited state ($n=2$) for $V=32^3$.
The cut-off dependence is investigated by varying the number of 
momenta $N \geq n$ for $N=0,1,2,3$.
%
%
\begin{figure}[t]
\vspace*{-0.6cm}
\centerline{\epsfxsize=7.0cm \epsfbox{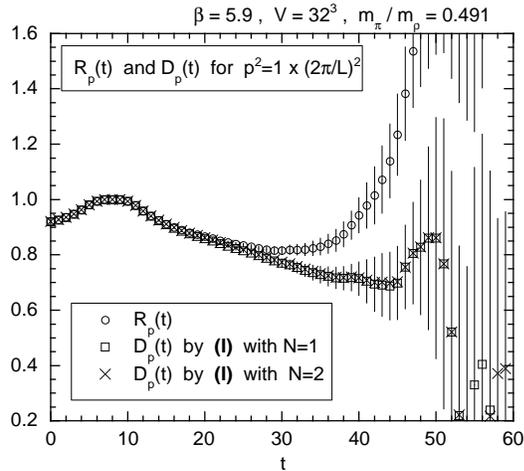}}
\vspace*{-1.0cm}
\caption{\label{Fig2.fig}
$R_p(t)$ and $D_p(t)$ for $p^2=(2\pi/L)^2$.
}
\vspace*{-0.6cm}
\end{figure}
%
%
\vspace{-0.2cm}
\section{ Results }
In order to examine the effects of diagonalization,
we calculate two ratios defined by
$R_{p}(t) \equiv G_{pp}(t) / [ G_{p}^\pi (t) ]^2$
and 
$D_{p}(t) \equiv \lambda_p (t) / [ G_p^\pi (t) ]^2$,
where $G_{p}^\pi (t)$ is the pion propagator with momentum $p$.
If $G_{pp}(t)$ or $\lambda_p(t)$ behaves as a single exponential function, 
we can obtain the energy shift $\Delta W_p$ easily 
from the ratio $R_p(t)$ or $D_p(t)$ by a single exponential fit.

In Fig.~\ref{Fig1.fig} we compare the ratios for $p^2=0$. The two-pion source
is placed at $t=8$.  
The momentum cut-off is set at $N=1$ and both diagonalizations are employed.
The signals are very clear and diagonalizations do not affect the result. 
We also checked the cut-off dependence by taking $N=2$ and confirmed that 
it is negligible.
In previous calculations of scattering lengths\cite{SGK,Kuramashi,GPS,JLQCD}
the ratio $R_0(t)$ was used to extract the energy shift $\Delta W_0$.
Our calculation demonstrates the reliability of these calculations.

We compare the ratios for $p^2=(2\pi/L)^2$ in Fig.~\ref{Fig2.fig}.
The momentum cutoff is set at $N=1$ and $N=2$. The method (I) is used.
In contrast to the $p^2=0$ case, the diagonalization is very effective. 
The cut-off dependence is negligible, however.
We also made the same analysis for method (II)
and confirmed that the results are independent of the method.

The analysis here, and additional one for $p^2=(2\pi/L)^2 \cdot 2$ 
for $V=32^3$, lead us to conclude that 
(i) The momentum cut-off should be taken $N \geq n$
for the energy shift $\Delta W_p$ ( $p^2=(2\pi/L)^2\cdot n$ ),  and 
%
%
(ii) both method (I) and (II) give same results.
%
%
\begin{figure}[t]
\vspace*{-0.6cm}
\centerline{\epsfxsize=7.0cm \epsfbox{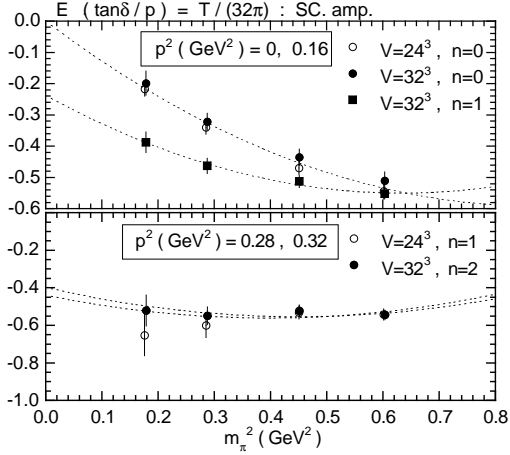}}
\vspace*{-1.0cm}
\caption{\label{Fig3.fig}
Results for scattering amplitude.
}
\vspace*{-1.0cm}
\end{figure}
%
%

In Fig.~\ref{Fig3.fig}
we plot our results of the scattering amplitude
$T/(32\pi)=E_p \cdot (\tan \delta (p) / p )$
obtained by substituting our data for $\Delta W_p$ into (\ref{Lusher.eq}).
The signals are clear in the small momentum region (upper figure in 
Fig.~\ref{Fig3.fig}) but become noiser for larger momenta (lower figure).
In order to obtain the scattering length and phase shift
for various momenta at the physical pion mass, we extrapolate our data
with the following fitting assumption :
$T/(32\pi)= A_{10} \cdot (m_\pi^2) + A_{20} \cdot (m_\pi^2)^2
+ A_{01} \cdot (p^2) + A_{02} \cdot (p^2)^2         
+ A_{11} \cdot (m_\pi^2 p^2)$.
The fit curves are also plotted in Fig.~\ref{Fig3.fig}.

\begin{figure}[t]
\vspace*{-0.6cm}
\centerline{\epsfxsize=7.0cm \epsfbox{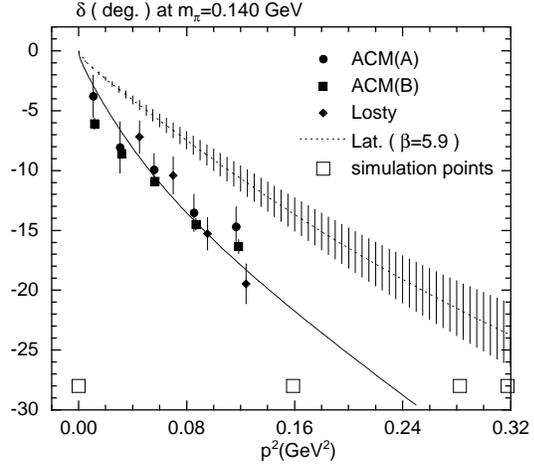}}
\vspace*{-1.0cm}
\caption{\label{Fig4.fig}
Comparison of our results of scattering phase shift $\delta(p)$ and 
experiments.
}
\vspace*{-0.6cm}
\end{figure}

In Fig.~\ref{Fig4.fig}
we compare our results of the phase shift $\delta(p)$
at physical pion mass obtained with the fit above 
with experiments~\cite{ACM_expt,Losty_expt}.
The simulation points are plotted by the large square symbols.
Our results for $\delta (p)$ are 40\% smaller in magnitude than those of 
experiments,  
and our result of scattering length, $a_0 m_\pi = -0.0266(16)$, 
differs from the ChPT prediction given by  
$ a_0 m_\pi = -0.044$ \cite{Gasser-Leutwyler:Bijinens}.

A possible cause of the discrepancy is finite lattice spacing effects. 
The JLQCD results for scattering length \cite{JLQCD} show sizable scaling 
violation, and hence that of the scattering phase shift cannot be 
considered small.
Further calculations nearer to the continuum limit or calculations with 
improved actions are desirable.
%
%
%
%
\hfill\break

This work is supported in part by Grants-in-Aid of the Ministry of Education 
(Nos.~10640246, 
10640248, 
11640250, 
11640294, 
12014202, 
12304011, 
12640253, 
12740133, 
13640260  
). 
VL is supported by the Research for Future Program of JSPS
(No. JSPS-RFTF 97P01102).
Simulations were performed on the parallel computer CP-PACS. 
%
%

%
%
\end{document}